\documentclass[aps,
twocolumn,
]{revtex4} 
\usepackage{epsf}
\usepackage{amsmath}
\usepackage{hyperref}
\usepackage{graphicx}
\newcommand*\euler{\mathrm{e}}
\begin{document}
\title{Conformational Mechanics of Polymer Adsorption Transitions\\ at Attractive Substrates}
\author{Monika M\"oddel}
\email[E-mail: ]{Monika.Moeddel@itp.uni-leipzig.de}
%
\author{Michael Bachmann}
\email[E-mail: ]{Michael.Bachmann@itp.uni-leipzig.de}
%
\author{Wolfhard Janke}
\email[E-mail: ]{Wolfhard.Janke@itp.uni-leipzig.de}
\homepage[\\ Homepage: ]{http://www.physik.uni-leipzig.de/CQT.html}
\affiliation{Institut f\"ur Theoretische Physik,
Universit\"at Leipzig, Postfach 100\,920, D-04009 Leipzig,\\
and Centre for Theoretical Sciences (NTZ), Emil-Fuchs-Stra{\ss}e 1, D-04105 Leipzig, Germany}
\begin{abstract}
Conformational phases of a semiflexible off-lattice homopolymer model near an attractive substrate are 
investigated by means of multicanonical computer simulations. In our polymer-substrate model, nonbonded 
pairs of monomers as well as monomers and the substrate interact via attractive van der Waals forces. To
characterize conformational phases of this hybrid system, we analyze thermal fluctuations of energetic and 
structural quantities, as well as adequate docking parameters. Introducing a solvent parameter related to the
strength of the surface attraction, we construct and discuss the solubility-temperature phase diagram. 
Apart from the main phases of adsorbed and desorbed conformations, we identify several other phase transitions 
such as the freezing transition between energy-dominated crystalline low-temperature structures and globular 
entropy-dominated conformations.
\end{abstract}
\maketitle
\section{\label{sec:intro}Introduction}
The study of the conformational behavior of polymers near 
surfaces is a fascinating field, both from a physical and chemical
perspective. It provides a rewarding playground for basic and 
applied research. With the advent of new sophisticated experimental 
techniques offering an enormous potential in polymer 
and surface research, the interest in the hybrid interface of 
organic and inorganic matter has increased. Among such
techniques at the nanometer scale is, for example, atomic force 
microscopy (AFM), where it is possible to measure the contour 
length and the end-to-end distance of individual polymers~\cite{gaub} 
or to quantitatively investigate the peptide adhesion on semiconductor 
surfaces~\cite{goede}. Another experimental tool with an extraordinary 
resolution in positioning and accuracy in force measurements 
are optical tweezers~\cite{smith,kremer}.

Applications for adsorption phenomena in polymeric solutions 
can be found in such different fields as lubrication, adhesion 
and surface protection, steric stabilization of colloidal particles,
as well as in biological processes of membrane-polymer interaction.
The understanding of the latter is particularly
important for the reconstruction of cell processes. The knowledge
of structure formation processes near interfaces is also a 
prerequisite for designing micro- or nanostructures providing a
large variety of possible applications in nanotechnology.

Despite much efforts in the past, the solvent-quality-dependent
behavior of a dilute polymer solution exposed to an adsorbing 
substrate is not yet fully understood. In good solvent, dominating 
structures are random coils since the monomers and the solvent 
molecules attract each other and, consequently, solvent molecules 
accumulate between monomers and push the monomers 
apart. Also at high temperatures, random coils are favored as
they possess a higher conformational entropy than globular 
conformations. Reducing the temperature, the more compact low-energy 
conformations gain thermodyna\-mic weight, and the 
polymer collapses in a cooperative rearrangement of the 
monomers. At the $\theta$-temperature, where the self-avoidance effect 
and the solvent effect exactly cancel, the size of a flexible 
polymer scales like an ideal chain, that is, $\left\langle R_{\rm gyr}^{2}\right\rangle \propto N^{2\nu}$, with $\nu=1/2$. 
Globular conformations are highly compact but have
only little internal structure. Hence those conformations are still 
entropy-dominated, and a further transition toward low-degenerate 
crystalline energetic states is expected and indeed observed: 
the freezing transition~\cite{wpaul,thomas}. 

The presence of an attractive surface strongly affects the 
behavior of the polymer in the vicinity of the interface. The
monomer-monomer attraction, being responsible for the collapse 
below the $\theta$-point, and the surface-monomer attraction, 
resulting in the adsorption, compete with each other. This 
competition gives rise to a variety of different conformational
phases. The polymer adsorbs at the surface, if the temperature 
is decreased below the adsorption transition temperature, but 
at high temperatures only a small number of monomers is in 
contact with the surface, even if the polymer was grafted to 
it~\citep{prellberg,jutta}. This is due to the lower entropy of conformations spread 
out on the surface, compared to the behavior in bulk.

Numerous detailed studies have been performed to elucidate 
the conformational behavior of homopolymers and heteropolymers
near substrates. Compared to experiments, computer 
simulations have the advantage that combinations of parameters 
can be varied at will. Theoretical studies have, for example, 
been performed analytically using scaling theory~\cite{eisenriegler,Usatenko}, mean-field 
density functional theory~\cite{wetting}, and series expansion~\cite{kumar,eisenriegler1} and 
numerically by employing off-lattice models such as a 
bead-spring model of a single polymer chain grafted to a 
weakly attractive surface~\cite{metzger,jutta}, multiscale modelling~\citep{luigi}, Monte Carlo 
studies of lattice homopolymers~\cite{prellberg,jutta,eisenriegler,michael,gupta,mbhetero,Whittington,grassberger,michaelhetero}, 
molecular dynamics 
combined with a stretching of an adsorbed homopolymer~\cite{Celestini}, or 
exact enumeration~\citep{kumar1}. Also adsorption-desorption dynamics were 
investigated in Brownian dynamics simulations of coarse-grained models~\cite{per}. \\

In this study, we performed multicanonical Monte Carlo
computer simulations in order to analyze thermodynamic
properties of the adsorption of a semiflexible polymer at a flat
and unstructured, attractive substrate. Our main objective is the
classification of the structural phases accompanying the adsorption
process and the construction of the complete (pseudo)phase
diagram parametrized by the temperature and a suitably
introduced solvent-quality parameter. The rest of the paper is 
organized as follows. In Sect.~\ref{sec:models}, the hybrid polymer-substrate
model, the multicanonical simulation method, as well as the 
measured observables are introduced. The main result, the 
pseudophase diagram, is presented and discussed in detail in 
Sect.~\ref{sec:results}. Several aspects of the phase structure are consolidated 
by a precise analysis of individual observables introduced
in Sect.~\ref{sec:models}. In Sect.~\ref{sec:lattice}, our off-lattice results are compared 
with former results obtained in simulations of lattice models.
Eventually, in Sect.~\ref{sec:summary}, the paper is concluded by a summary of our findings. 

\section{\label{sec:models}The Model}
\subsection{Hybrid Modeling of Polymer-Substrate Interaction.}
We employ a coarse-grained off-lattice model for semiflexible
homopolymers that has also been generalized for studies of
heteropolymers~\cite{Still93,Still95} and helped to understand protein folding
channels from a mesoscopic perspective~\cite{stefan1,stefan2}. In contrast to earlier
adsorption studies of lattice polymers~\cite{prellberg,jutta,eisenriegler,michael,gupta,mbhetero,Whittington,grassberger,michaelhetero}, we here accept
the associa\-ted additional computational cost of an off-lattice 
model in order to get rid of undesired effects of underlying 
lattice symmetries. 

As on the lattice, we assume that adjacent monomers are
connected by rigid covalent bonds. Thus, the distance $|\vec{r}_{i+1}-\vec{r}_{i}|$ is 
fixed and set to unity. Bond and torsional angles are free to
rotate. The energy function consists of three terms,
\begin{equation}
 E=E_{\rm bend}+E_{\rm LJ}+E_{\rm sur}
\end{equation} 
associated with the bending stiffness ($E_{\rm bend}$), monomer-monomer 
Lennard-Jones interaction ($E_{\rm LJ}$), and monomer-surface attraction 
($E_{\rm sur}$).
 
\begin{figure}
\includegraphics[width=8.cm]{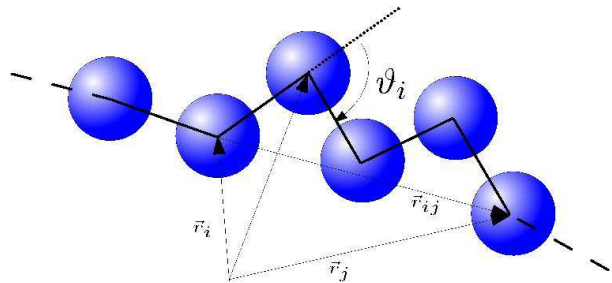}
\caption{\label{fig:1} A segment of the coarse-grained semiflexible polymer. The 
distance between two adjacent monomers is fixed and set to unity. The 
bending angle at the $\left(i+1\right)$th monomer is denoted by $\vartheta_{i}$, and the vector between 
the $i$th and $j$th monomer by $\vec{r}_{ij}\equiv\vec{r}_{j}-\vec{r}_{i}$ with $|\vec{r}_{ij}|={r}_{ij}$.}
\end{figure}

A sketch of a coarse-grained polymer segment is depicted in
Figure \ref{fig:1}.
The position vector of the $i$th monomer, $i=1,\ldots, N$, 
is denoted by $\vec{r}_{i}$. A polymer with $N$ monomers has $N-1$ bonds 
of length unity between neighboring monomers and $N-2$ 
bending angles $\vartheta_{i}$, $i=1,\ldots, N-2$, defined through
\begin{equation}
 \cos \left(\vartheta_{i} \right) 
=\left( \vec{r}_{i+1}-\vec{r}_{i}\right)\cdot\left( \vec{r}_{i+2}-\vec{r}_{i+1}\right) .
\end{equation} 
The Lennard-Jones potential of nonbonded monomers is of
standard form,
\begin{equation}
 E_{\rm LJ}=4\sum_{i=1}^{N-2}\sum_{j=i+2}^{N}\left( \dfrac{1}{r_{ij}^{12}}-\dfrac{1}{r_{ij}^{6}}\right) ,
\end{equation} 
where ${r}_{ij}=|\vec{r}_{j}-\vec{r}_{i}|$. The lowest-energy distance of the 
Lennard-Jones potential between two monomers is $r_{\rm min}=\sqrt[6]{2}\approx 1.12$ 
and is hence slightly larger than the unity bond length.
The bending energy is defined as 
\begin{equation}
 E_{\rm bend}=\dfrac{1}{4}\sum_{i=1}^{N-2}\left( 1-\cos\left(\vartheta_{i} \right) \right) .
\end{equation} 
The angle $\vartheta_{i}$ is in the interval $\left[ 0,\pi\right) $ and the bending energy 
can be considered as a penalty for successive bonds deviating 
from a straight arrangement. 

The attractive surface potential is obtained by integrating over 
a smooth half-space, where every space ele\-ment interacts with 
a single monomer by the usual Lennard-Jones 12-6 expression. 
One obtains the potential~\cite{steelesurface,hentschke}:
\begin{equation}
 E_{\rm sur}= \epsilon_{s}\sum_{i=1}^{N}\left[\dfrac{2}{15}\left( \dfrac{1}{z_{i}}\right)^{9} - \left( \dfrac{1}{z_{i}}\right)^{3}\right],
\label{eq:esur}\end{equation} 
where $z_i$ is the distance of the $i$th monomer from the surface. 
The parameter $\epsilon_s$ defines the surface attraction strength, and as 
such it weighs the energy scales of intrinsic monomer-monomer 
attraction and monomer-surface attraction.

A distance $z=L_{\rm box}$ away from the attractive surface, we place 
a steric wall that is ne\-cessary to prevent the molecule from 
escaping. Because the exact form of the density of states depends 
on the box height $L_{\rm box}$, all measured quantities also depend on 
the choice of $L_{\rm box}$. As soon as the box size exceeds the polymer 
size, however, the influence on the observables, apart from the 
substrate distance of the center-of-mass $z_{\rm cm}=\sum_{i=1}^{N}z_i/N$ of the 
polymer, is sufficiently small. We chose $L_{\rm box}=20\,(40)$ for the 
polymer with $N=13\,(20)$ monomers.
\subsection{Multicanonical Method.}
The canonical partition function
of our hybrid system at temperature $T$ is given in natural units
by
\begin{equation}
Z=\int_{E_{\rm min}}^{\infty} \! {\rm d}\!E\, g(E) \euler^{-E/T},
\label{eq:partitionsum}
\end{equation} 
where $g(E)=\euler^{S(E)}$ is the density of states that connects
(microcanonical) entropy $S$ and energy. Therefore, all information
regarding the phase behavior of the system -- typically
governed by the competition between entropy and energy -- is
encoded in $g(E)$. Consequently, for a detailed global analysis
of the phase behavior, a precisely estimated density of states is
extremely helpful. Unfortunately, the density of states covers
many orders of magnitude in the phase transition regimes, so
that its estimation requires the application of sophisticated
generalized-ensemble Monte-Carlo methods.

For our analyses, we have performed multicanonical
simulations~\cite{bergneuhaus}, as multicanonical sampling allows the estimation
of $g(E)$, in principle, within a single simulation. The
idea of the multicanonical method is to increase the sampling
rate of conformations being little favored in the free-energy
landscape and, finally, to perform a random walk in energy
space. This is achieved in the simplest way by setting $T=\infty$ 
and introducing suitable multicanonical weights $W_{\rm muca}(E)\sim g^{-1}(E)$
in order to sample conformations $\bf X$ according to a 
transition probability
\begin{equation}
\omega(\mathbf{X}\rightarrow \mathbf{X'})=\min [\euler^{S(E(\mathbf{X}))-S(E(\mathbf{X'}))},1],
\label{eq:transition}
\end{equation} 
where $S(E(\mathbf{X}))=-\ln W_{\rm muca}(E(\mathbf{X}))=\ln g(E(\mathbf{X}))$.

The implementation of multicanonical sampling is not
straightforward as the multicanonical weights $W_{\rm muca}(E)$ are
obviously unknown a priori. Therefore, starting with $W_{\rm muca}^{0}(E)={\rm const.}$,
the weights have to be determinded by an iterative
procedure until the multicanonical histogram is almost ''flat'',
that is, if the estimate for the density of states after the $n$th run,
${\hat{g}}^{(n)}(E)$, satisfies
\begin{equation}
{\hat{g}}^{(n)}(E) W_{\rm muca}^{n-1}(E)\approx {\rm const.}
\label{eq:constdos}
\end{equation} 
in the desired range of energies. An efficient, error-weighted
estimation method for the multicanonical weights is described
in detail in refs \cite{janke} and \cite{berg1}.

Eventually, if eq \ref{eq:constdos} is reasonably satisfied, the multicanonical
weights $W_{\rm muca}^{n}(E)=[\hat{g}^{(n)}(E)]^{-1}$ are then used in a final long
production run, where all quantities of interest are measured
and stored in a time-series file. The canonical expectation value
of any quantity $O$ at temperature $T$ is then obtained from the
multicanonical time series of length $M$ by reweighting,
\begin{equation}
\langle O\rangle =\frac{\sum_{i=1}^M O(\mathbf{X}_t) \euler^{-E(\mathbf{X}_t)/T} W_{\rm muca}^{-1}(E(\mathbf{X}_t))}    
 {\sum_{i=1}^M  \euler^{-E(\mathbf{X}_t)/T} W_{\rm muca}^{-1}(E(\mathbf{X}_t))},
\label{eq:canexp}
\end{equation} 
where $t$ is the multicanonical Monte Carlo ''time'' step (or sweep).

\subsection{Suitable Energetic and Structural Quantities for Phase Characterization.}
To obtain as much information as possible 
about the canonical equilibrium behavior, we define the following
suitable quantities $O$. Next to the canonical expectation 
values $\left\langle O\right\rangle$, we also determine the fluctuations about these
averages, as represented by the temperature derivative $\left( \left\langle O E \right\rangle -\left\langle O\right\rangle \left\langle E\right\rangle \right) /T^2$.
We use generic units, in which $k_B=1$.

Apart from energetic fluctuations such as the specific heat, 
$c_{V}={\rm d}\left\langle E/N\right\rangle{\rm d}T$, several structural quantities are of interest. The 
radius of gyration is a measure for the extension of the polymer 
and defined by $R_{\rm gyr}^{2}\equiv\sum_{i=1}^{N}\langle ( \vec{r}_i-\vec{R}_{\rm cm}) ^{2}\rangle/N=
\sum_{i=1}^{N}\sum_{j=1}^{N}\langle ( \vec{r}_i-\vec{r}_{j}) ^{2}\rangle /2N^{2}$ with 
$\vec{R}_{\rm cm}=\sum_{i=1}^{N}\vec{r}_i/N$ being the center-of-mass of the polymer. 
Since the substrate introduces a 
structural anisotropy into the system, it is not only useful to 
investigate the overall compactness of the polymer expressed 
by $\left\langle R_{\rm gyr}\right\rangle $, but also to study the expected different behavior of 
its components parallel and perpendicular to the surface:
 $R_{\parallel}^{2}=\sum_{i=1}^{N}\sum_{j=1}^{N}\langle ( x_i-x_{j}) ^{2}+( y_i-y_{j}) ^{2}\rangle /2N^{2}$
and
$ R_{\perp}^{2}=\sum_{i=1}^{N}\sum_{j=1}^{N}\langle ( z_i-z_{j}) ^{2}\rangle /2N^{2} $
such that $R_{\rm gyr}^{2}=R_{\parallel}^{2}+R_{\perp}^{2}$. 
\begin{figure}
	\includegraphics[width=8.8cm]{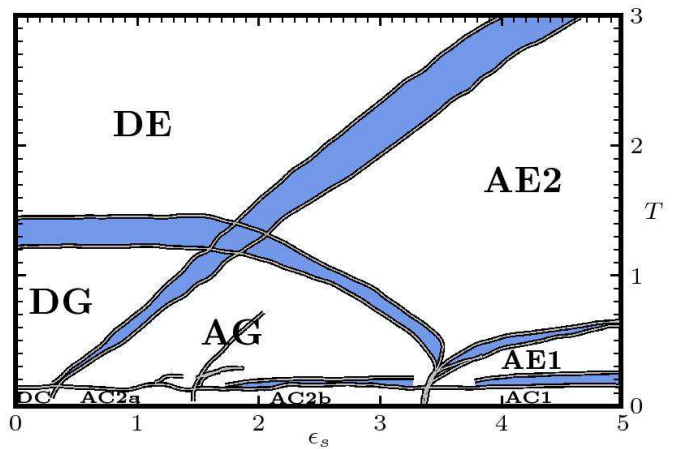}
	\caption{\label{fig:4} Phase diagram of the 20mer. The colored stripes separate 
the individual conformational phases (see text and Figure \ref{fig:5}).}
\end{figure}
\begin{figure}
	\includegraphics[width=7.8cm]{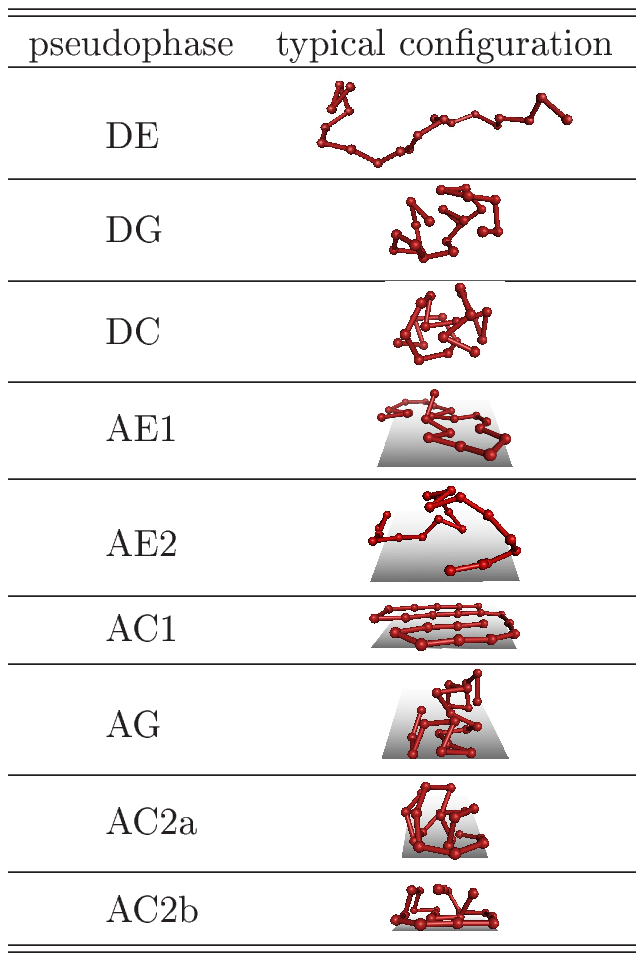}
	\caption{\label{fig:5} Representative examples of conformations for the 20mer in
the different regions of the $T$-$\epsilon_s$ pseudophase diagram. DE, DG, and 
DC represent bulk ''phases'', where the polymer in preferably desorbed.
In regions AE1, AE2, AC1, AG, AC2a, and AC2b, conformations are favorably adsorbed.}
\end{figure}

Clear evidence that the polymer is, on average, freely moving 
in the box or very close to the surface can be provided by the 
distance of the center-of-mass, $z_{\rm cm}$, of the polymer to the surface.

Another useful quantity is the mean number of monomers 
docked to the surface. A single-layer structure is formed if all
monomers are attached at the substrate; if none is attached, the
polymer is quasifree (desorbed) in solvent. The surface potential 
is a continuous potential, and in order to distinguish monomers 
docked to the substrate from those not being docked, it is 
reasonable to introduce a cutoff. We define a monomer $i$ as 
being ``docked'' if $z_i<z_c\equiv 1.2$. The corresponding measured 
quantity is the average ratio $\left\langle n_s\right\rangle$ of monomers docked to the 
surface and the total number of monomers. This can be expressed as
$ n_s = {N_s}/{N}$ with $ N_s=\sum_{i=1}^N \Theta( z_c-z_i) $,
where $\Theta(z)$ is the Heaviside step function.        

Similary, the mean number of intrinsic contacts is introduced
as a measure of the global compactness of the polymer: $ n_m = {N_m}/{N}$ 
with $N_m=\sum_{i=1}^{N-2}\sum_{j=i+2}^N \Theta\left(e_c-e_{\rm LJ}\left( r_{ij}\right)  \right)$, where $e_c\equiv -0.2$
and $e_{\rm LJ}\left( r_{ij}\right) =4\left( r_{ij}^{-12}-r_{ij}^{-6}\right) $. 

\section{\label{sec:results}Results and discussion}
Our major objective is the construction of the pseudophase
diagram of conformational phases, based on our results obtained
for energetic and structural fluctuations. To this end, multicanonical
simulations~\citep{bergneuhaus} for 51 different surface attraction strengths $\epsilon_s$, in the
range $\epsilon_s=0,\ldots, 5$, were performed and reweighted to temperatures
$T\in(0,5]$. Each simulation consisted of $10^8$ sweeps and was 
performed with at least 2 different initializations of the random 
number generator in order to avoid systematic deviations. 

For a convenient overview, we display our final pseudophase 
diagram of the 20mer already here in Fig.~\ref{fig:4}. Conformations
representative of the different phases are shown in Fig.~\ref{fig:5}. 
The details will be discussed in the following. It should be 
stressed that all phases and transitions occuring in our analysis
are not phases in the strict thermodynamic sense, since we are 
dealing with finite chain lengths. However, even for the rather
short chains considered here, a reasonable picture of the polymer
adsorption behavior at the surface is obtained, and most of the 
phases are believed to still exist for longer chains.
\begin{figure}
 		\includegraphics[width=8.7cm]{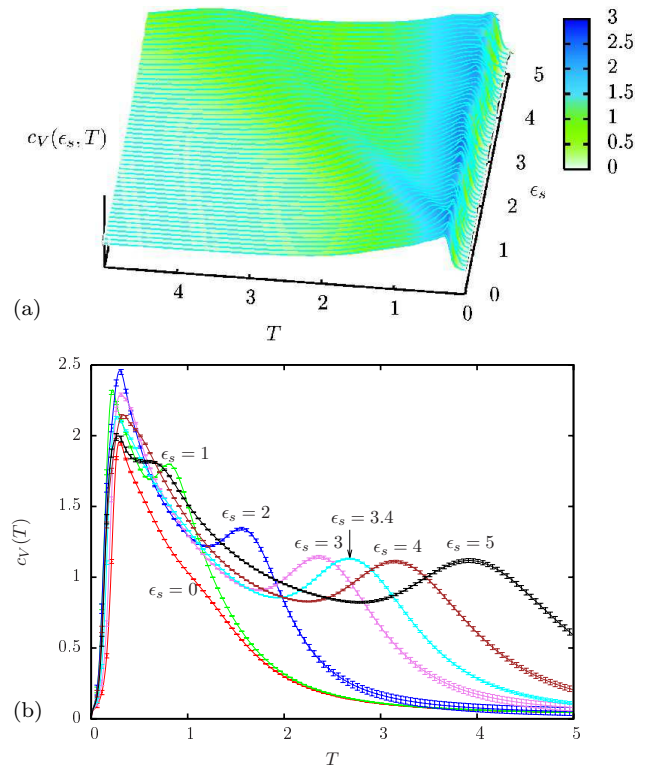}
\caption{\label{fig:EE3d} (a) Specific heat as a function of temperature $T$ and surface 
interaction strength $\epsilon_s$ for the 20mer. Lines represent the simulation 
data for fixed values of $\epsilon_s$, the color code is interpolated. (b) Specific 
heat curves for different values of $\epsilon_s$. }
\end{figure}
\subsection{Energetic Fluctuations}
In Fig.~\ref{fig:EE3d}(a), the specific heat 
of the 20mer is shown as a function of $T$ and $\epsilon_s$. Although for 
both investigated polymer lengths the energy varies smoothly 
with changing $T$ and $\epsilon_s$ (with a global minimum at mini\-mal 
temperature and maximal surface attraction), two transitions can 
be identified as ridges in the profile of the specific heat: The 
first one is the \textit{adsorption transition} separating desorbed and 
adsorbed conformations. The other transition is sort of a \textit{freezing transition} 
at low temperatures. Near $T= 0.25$, the specific heat 
exhibits a pronounced peak and decreases rapidly at lower 
temperatures, independently of the surface attraction strength. 
The low-temperature peak in specific heat and the crystalline 
shape of structures found in this regime signalize this freezing
transition. Although the freezing temperature seems to be rather 
constant, the type of crystalline structure adopted by the polymer 
depends strongly on $\epsilon_s$. For the identification of the polymer 
shapes, we take in the next subsection a closer look at the 
conformational quantities.
\subsection{Comparative Discussion of Structural Fluctuations}
\begin{figure}
 		\includegraphics[width=8.7cm]{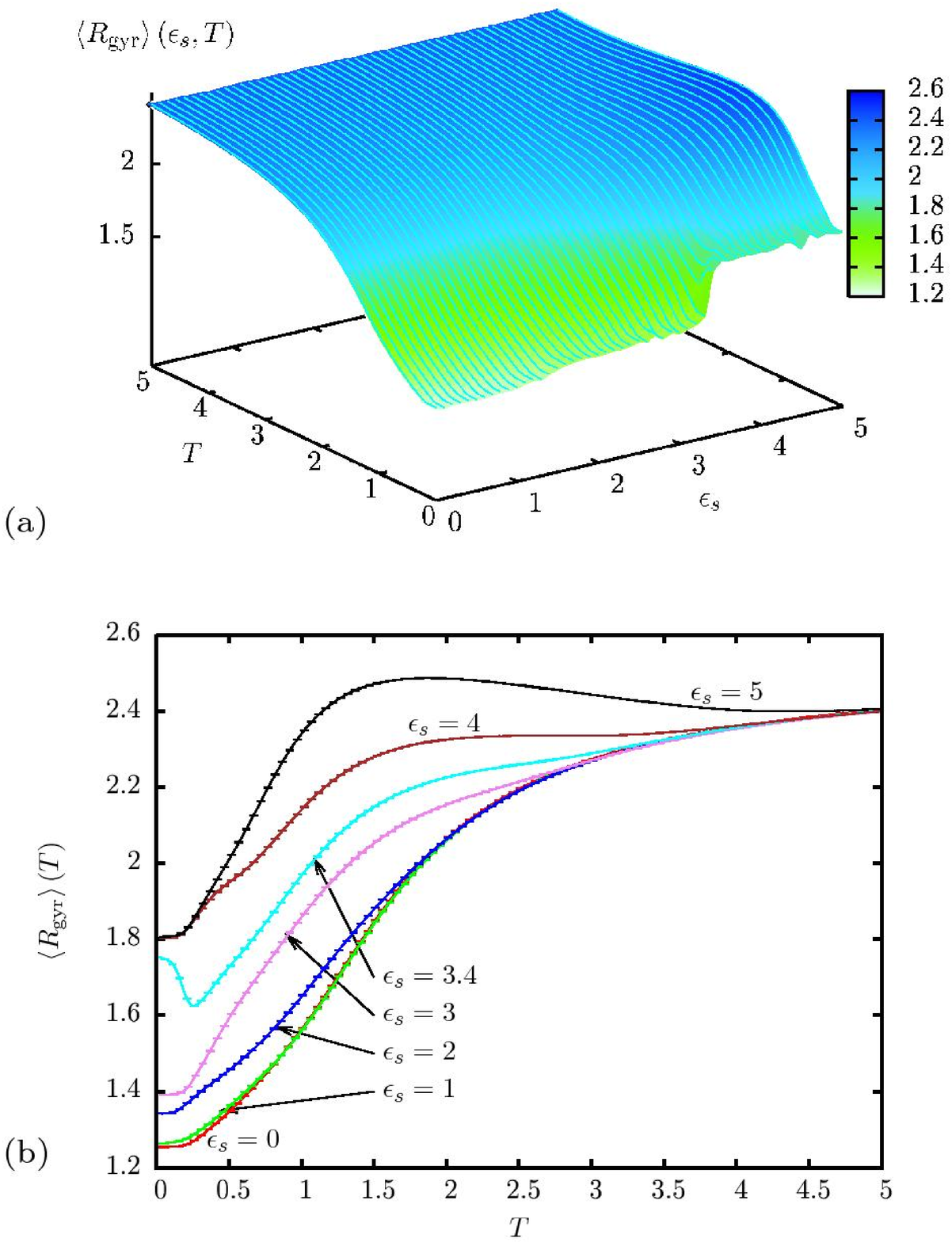}
\caption{\label{fig:Rgyr3d} (a) $\left\langle R_{\rm gyr} \right\rangle $ of the 20mer as a function of $T$ and $\epsilon_s$. (b) 
$\left\langle R_{\rm gyr} \right\rangle $ as a function of the temperature for various values of $\epsilon_s$.}
\end{figure}

\begin{figure}
 		\includegraphics[width=8.7cm]{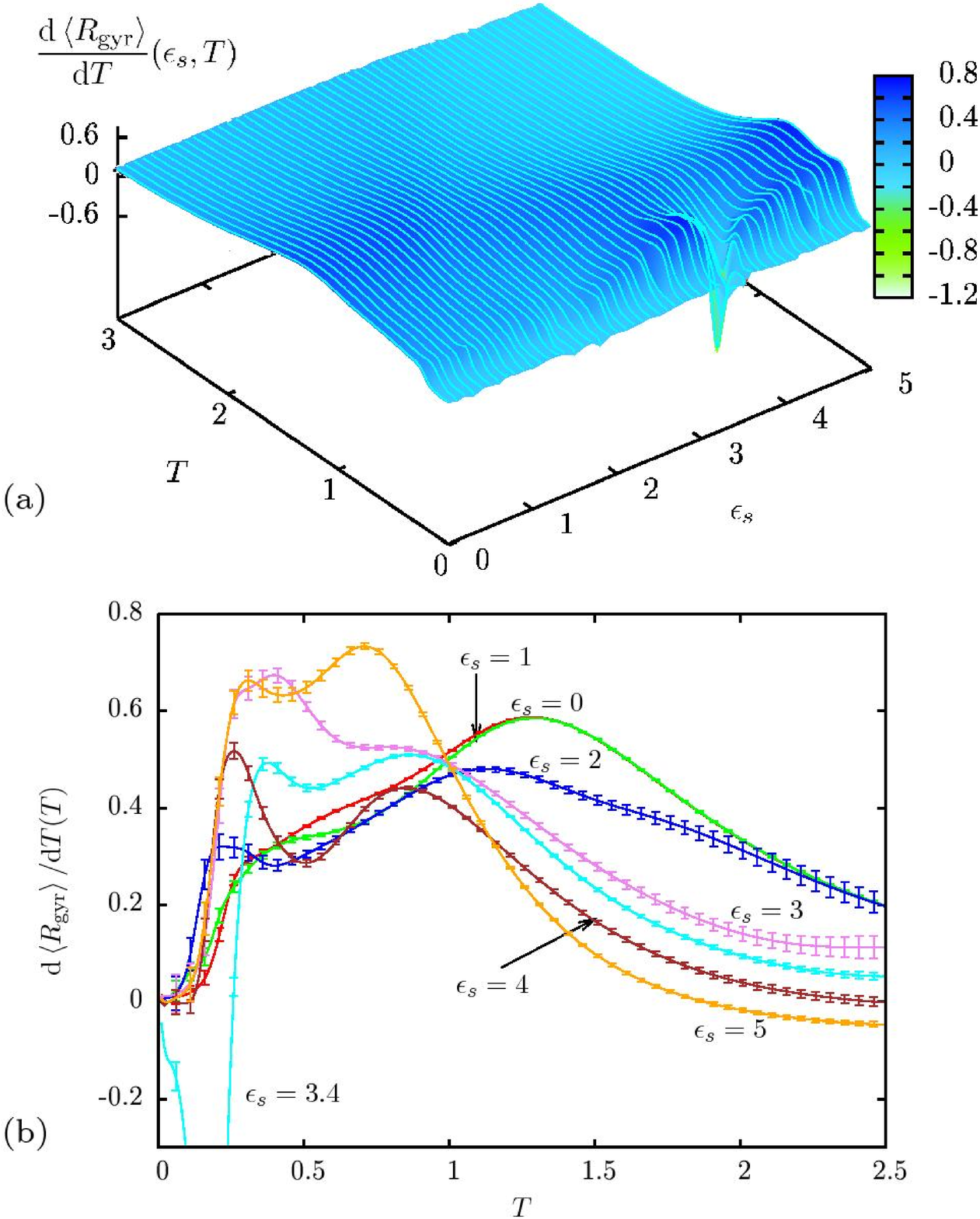}
\caption{\label{fig:RgyrE3d} (a) ${\rm d}\left<R_{{\rm gyr}}\right>/{\rm d}T$ of the 20mer. (b) ${\rm d}\left<R_{{\rm gyr}}\right>/{\rm d}T$, parametrized
by $\epsilon_s$. }
\end{figure}

The radius of gyration provides an excellent measure of the 
globular compactness of polymer conformations. Figures \ref{fig:Rgyr3d} and
\ref{fig:RgyrE3d} reveal that the most compact conformations dominate at low 
temperatures and for small values of surface attraction strengths 
$\epsilon_s$. The freezing transition can be found at the same temperature, 
as it has already been identified from the specific heat. The 
adsorption transition typically affects only segments of the 
polymer and is thus not prominently signaled by the radius of gyration. 

\begin{figure}
 		\includegraphics[width=8.7cm]{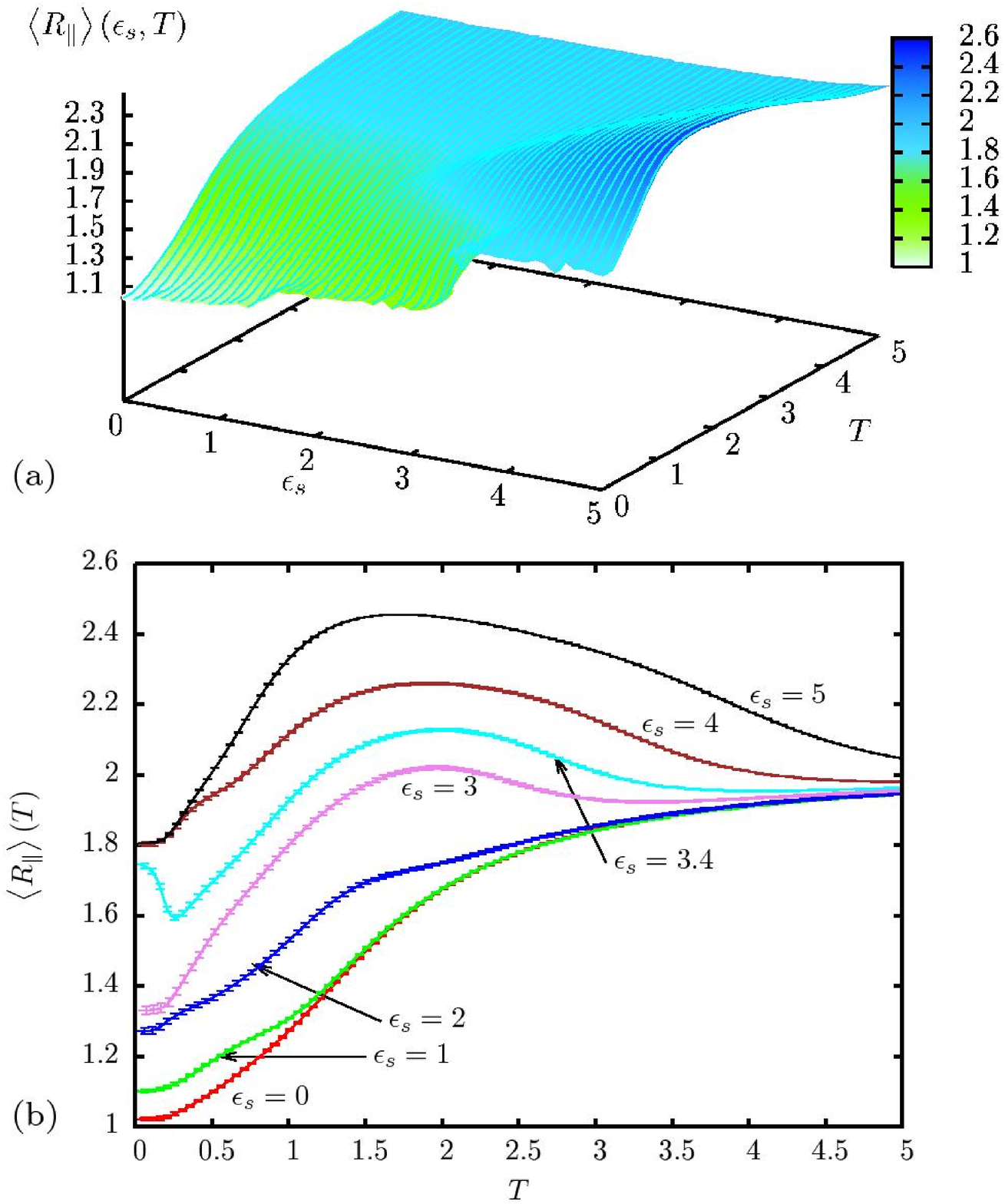}
\caption{\label{fig:Rgyrxy} (a) $\left\langle R_{\parallel} \right\rangle $ of the 20mer. (b) $\left\langle R_{\parallel} \right\rangle $ 
for selected values of $\epsilon_s$. }
\end{figure}

\begin{figure}
 		\includegraphics[width=8.7cm]{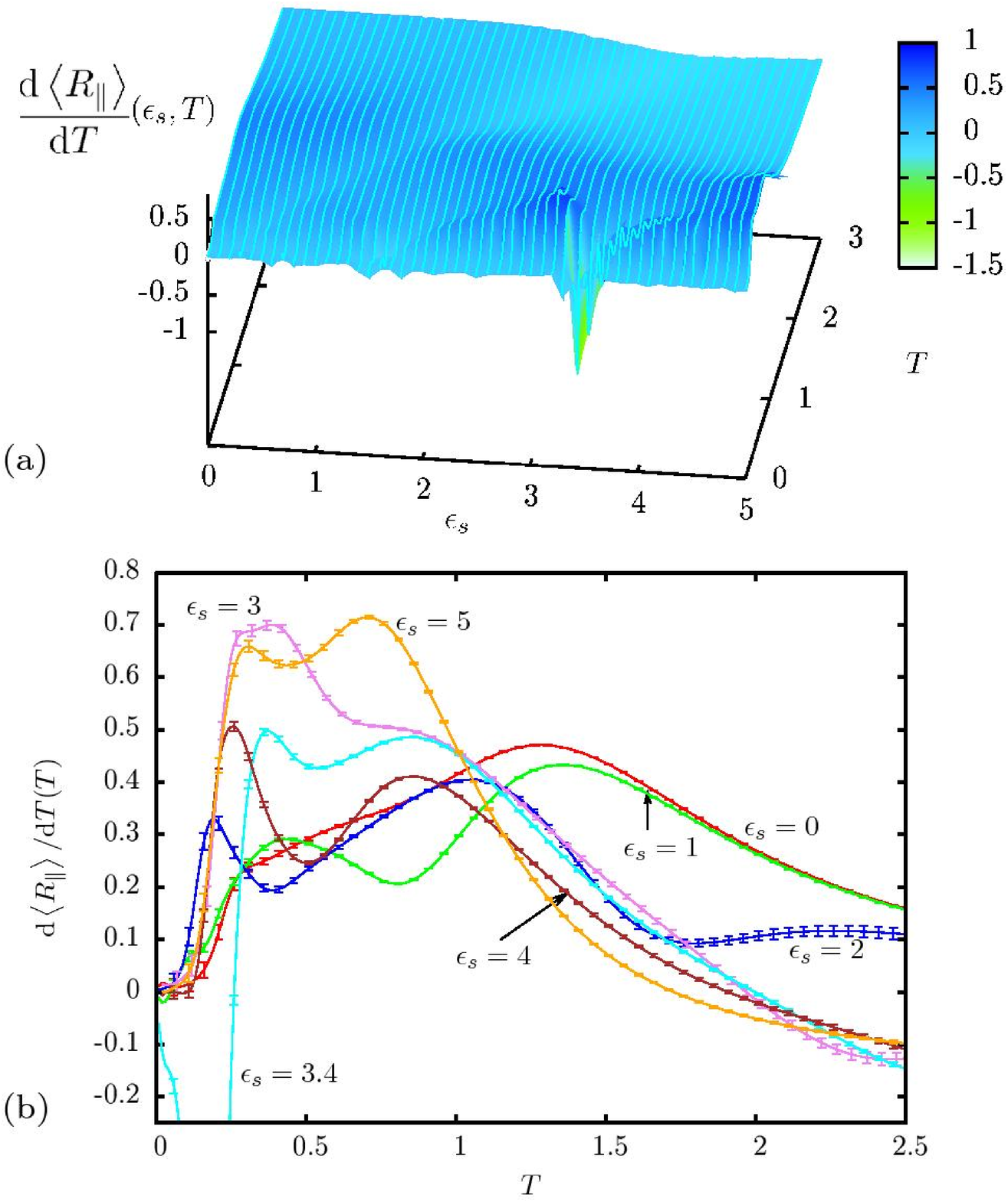}
\caption{\label{fig:RgyrxyE} (a) ${\rm d}\left<R_{\parallel}\right>/{\rm d}T$ for the 20mer. (b) ${\rm d}\left<R_{\parallel}\right>/{\rm d}T$ 
for different surface attraction strengths $\epsilon_s$. }
\end{figure}

\begin{figure}
 		\includegraphics[width=8.7cm]{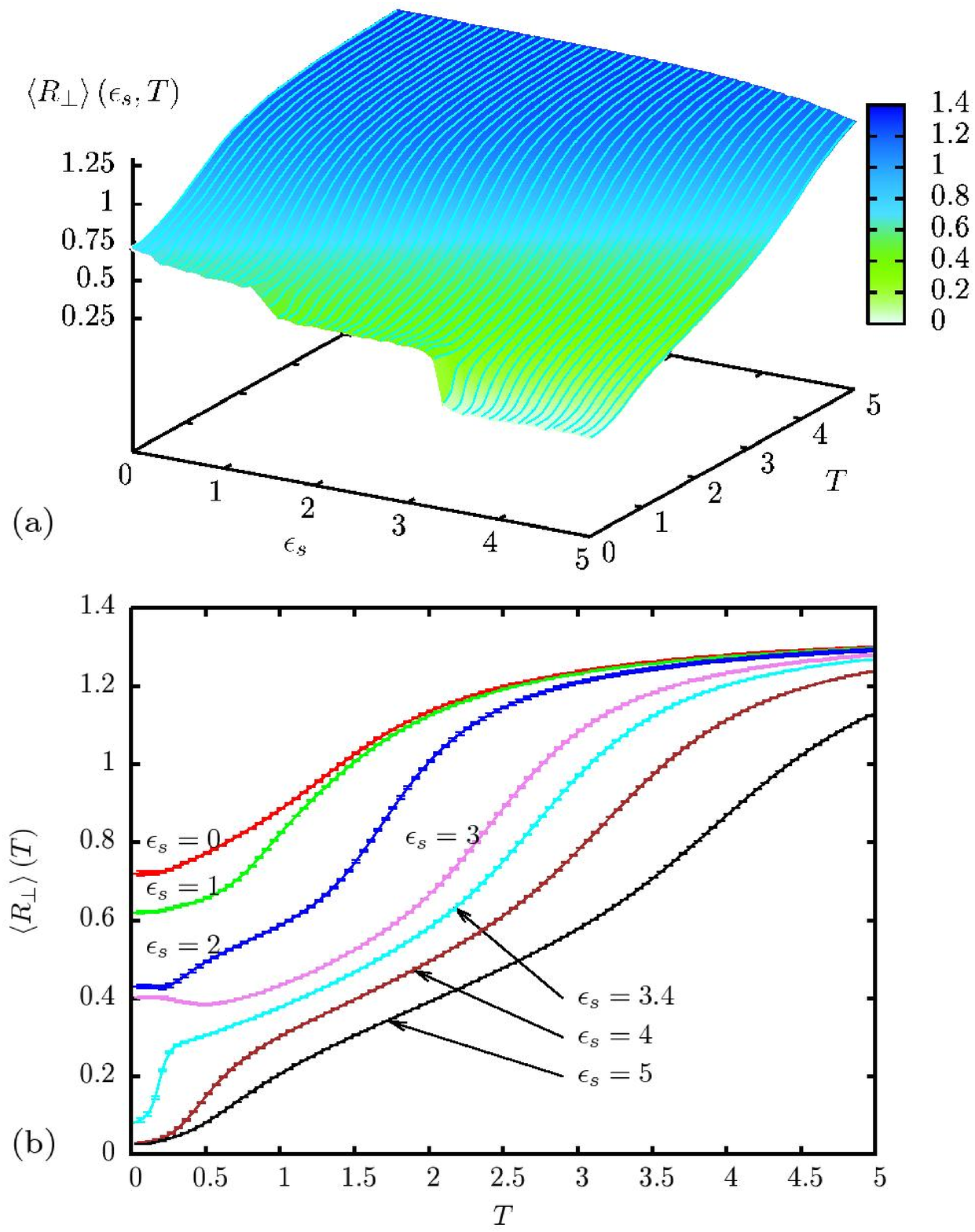}
\caption{\label{fig:Rgyrz} (a) $\left\langle R_{\perp} \right\rangle $ of the 20mer. (b) $\left\langle R_{\perp} \right\rangle $ for selected values of $\epsilon_s$.  }
\end{figure}

\begin{figure}
 		\includegraphics[width=8.7cm]{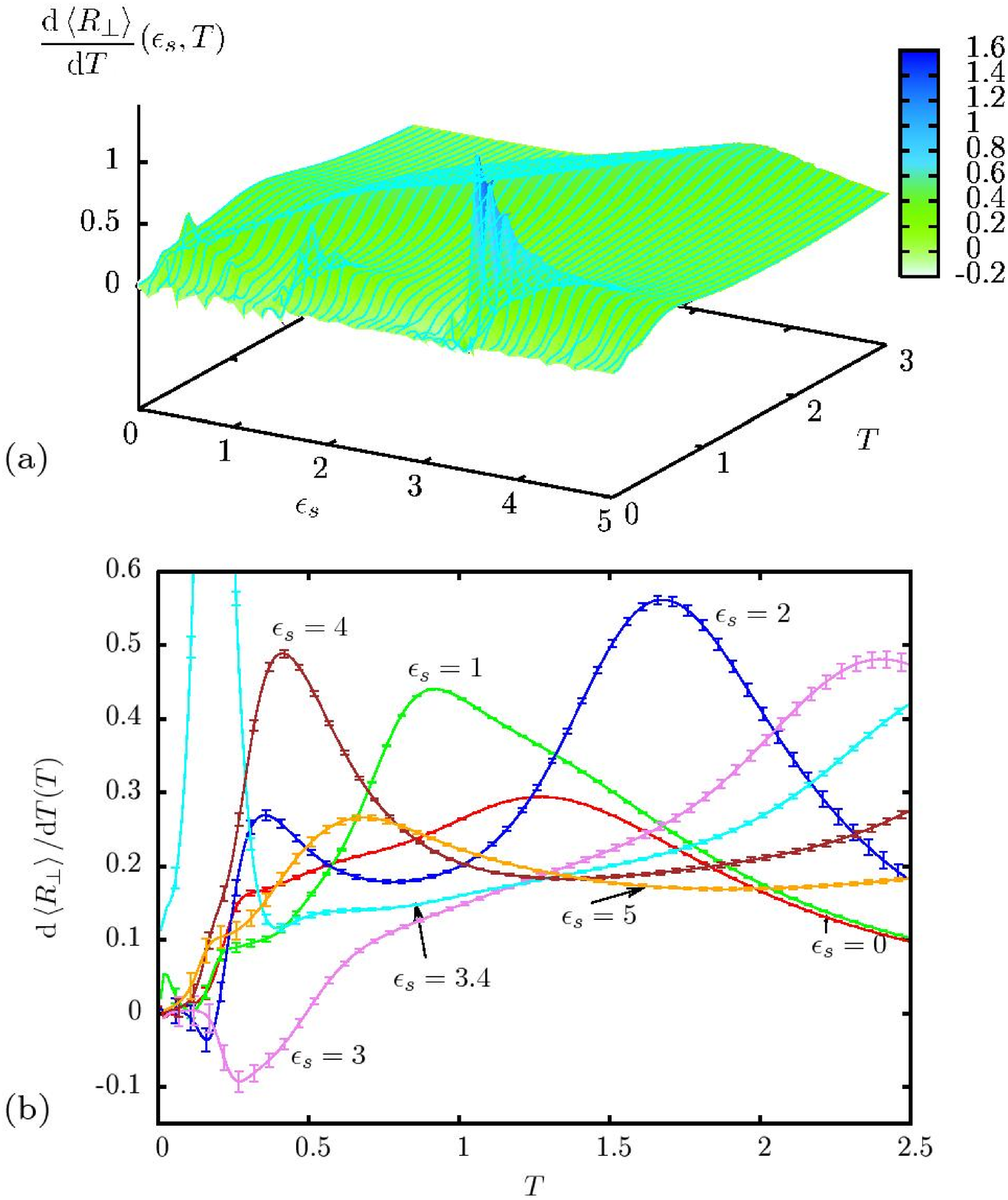}
\caption{\label{fig:RgyrzE} (a) ${\rm d}\left<R_{\perp}\right>/{\rm d}T$ of the 20mer. (b) ${\rm d}\left<R_{\perp}\right>/{\rm d}T$ for different 
surface attraction strengths $\epsilon_s$. }
\end{figure}

This is different for its components parallel (Figures \ref{fig:Rgyrxy} and 
\ref{fig:RgyrxyE}) and perpendicular (Figures \ref{fig:Rgyrz} and \ref{fig:RgyrzE}) to the surface, 
respectively. For example, for $\epsilon_s \geq3.4$, $\left\langle R_{\perp} \right\rangle $ vanishes at low 
temperatures for the 20mer, whereas $\left\langle R_{\parallel} \right\rangle $ attains low values at 
lower $\epsilon_s$. The vanishing of $\left\langle R_{\perp} \right\rangle $ corresponds to conformations
where the polymer is spread out flat on the surface without any 
extension into the third dimension. The associated pseudophases
are called adsorbed compact (AC1) and adsorbed expanded 
(AE1) phase. The `1' is appended in order to distinguish these
phases from topological three-dimensional phases, such as, 
for example, the AC2 subphases. The phases AC1 and AE1 
are separated by the freezing transition such that polymer
structures in AC1 are maximally compact at lower temperatures,
whereas AE1 conformations are less compact and more flexible
but still lie rather planar at the surface. 
To verify that conformations in AC1 are indeed maximally 
compact single layers, we argue as follows. The most compact 
shape in the two-dimensional (2D) continuous space is the 
circular disk. Thus, one can calculate $\left\langle R_{\parallel} \right\rangle $ for a disk and compare 
it with the simulated value. Assuming $N$ monomers to be 
distributed evenly in the disk, $N\approx\pi r^{2}$, where $r$ is the radius of 
the disk in units of the mean distance of neighboring monomers. 
The radius of gyration in the same units is thus given by
\begin{equation}
 {R_{\rm gyr,disk}}^{2}=\dfrac{1}{\pi r^{2}}\int_{r'\leq r} {\rm d}^{2}r'r'^{2}=\dfrac{1}{2}r^{2}\approx\dfrac{N}{2\pi}.
\end{equation}
Since we have two types of mean distances between monomers 
in compact conformations dependent on whether they are 
adjacent along the chain or not, we expect for disk-like
conformations an the surface: $\sqrt{20/2\pi}\approx1.784< \left\langle R_{\parallel,20} \right\rangle<2.003\approx r_{\rm min}\sqrt{20/2\pi}$. 
The simulated value is $\langle R_{\parallel,20}\rangle\approx 1.81$, 
which nicely fits the estimate. The equivalent estimate for the 
13mer also confirms discoidal conformations in AC1.

The argument is similar for sphere-like three-dimensional 
(3D) compact conformations with $N=4\pi r^{3}/3$. Corresponding 
conformations are found as free desorbed compact chains (DC), 
as well as adsorbed compact polymer conformations (AC2a) 
for weak surface attraction. In this case, the radius of gyration 
is given by
\begin{equation}
  {R_{\rm gyr,sph}}^{2}=\dfrac{1}{4\pi r^{3}/3}\int_{r'\leq r}{\rm d}^{3}r'r'^{2}=\dfrac{3}{5}r^{2}\approx\dfrac{3}{5}\left(\dfrac{3N}{4\pi} \right) ^{2/3}.
\end{equation}
The estimates $(3/5)^{1/2}(3\times 13/4\pi)^{1/3}\approx1.130<\langle R_{{\rm gyr},13}\rangle<1.268\approx r_{\rm min}(3/5)^{1/2}(3\times 13/4\pi)^{1/3}$, 
and $1.684<\langle R_{{\rm gyr},20}\rangle<1.464$ slightly overestimate the simulated values $\langle R_{{\rm gyr},13}\rangle=1.023$ and $\langle R_{{\rm gyr},20\rangle}=1.242$.
The deviations can be explained by the fact that the 
mass of the polymer is not uniformly distributed in the sphere 
as it is assumed in the calculation. For a compact packing of 
discrete monomer positions, it is more realistic that the outer 
thin shell of the sphere does not contain any monomers. 
Performing the integration not from $r'=0$ to $r'=r$, but only 
to $r'=r-\varepsilon$, reduces the estimated radius of gyration 
significantly already for small $\varepsilon$ due to the increased weight of 
the outer shells in higher dimensions. Taking this effect into 
account, the thus obtained values of $\left\langle R_{{\rm gyr}}\right\rangle $ seem to be even more reasonable. 

The most pronounced transition is the strong layering 
transition at $\epsilon_s\approx3.4$ for $N=20$ that separates regions of planar 
conformations (AC1, AE1) in the $T$-$\epsilon_s$ plane from the region 
of stable double-layer structures (AC2b) and adsorbed globules 
(AG), below and above the freezing transition, respectively. For
high surface attraction strengths $\epsilon_s$, it is energetically favorable 
to form as many surface contacts as possible. In the layering-transition
region, a higher number of monomer-monomer 
contacts causes the double-layer structures to have just the same 
energy as single-layer structures. For lower $\epsilon_s$ values, the double-layer 
structures possess the lowest energies. Hence this transition is a sharp energe\-tical transition. 

Although for the considered short chains no higher-layer
structures are observed, the components $\left\langle R_{\parallel,\perp} \right\rangle $ indicate some 
activity for low surface attraction strengths. For $N=20$, $\epsilon_s\approx1.4$ 
is the lowest attraction strength, where still stable double-layer 
conformations are found below the freezing transition. 
What follows for lower  $\epsilon_s$ values after a seemingly continuous 
transition is a low-temperature subphase of surface attached 
compact conformations, called AC2a. AC2a conformations 
occur if the monomer-surface attraction is not strong enough 
to induce a layering in compact attached structures. The characterization
of structures in this subphase requires some care, as
system-size effects are dominant. Although the surface attraction
is sufficiently strong to enable polymer-substrate contacts,
compact desorbed polymer conformations below the $\theta$-transition 
are not expected to change much. Thus, layering effects do not 
occur. We found two distinct classes of structures in this region:
(1) completely undistorted compact conformations touching the
surface and (2) semispherically shaped structures docked at the 
surface. For the rather short chains in our study, the occurrence
of these shapes strongly depends on the precise number of
monomers. Not surprisingly, we find differences for the 13mer
and the 20mer. 

For $N=13$, both components of the radius of gyration, ${\rm d}\left<R_{\perp}\right>/{\rm d}T$ 
and ${\rm d}\left<R_{\parallel}\right>/{\rm d}T$, indicate a transition at $\epsilon_s\approx0.45$. Low-energy 
configurations reveal that this is a wetting transition between 
undistorted compact conformations for smaller $\epsilon_s$ and docked 
conformations for largerr surface attraction strengths. An analogous 
transition for $N=20$ was not found. In this case, the AC2a 
pseudophase seems to consist of a mixture of both types without 
any transition between them. This is also confirmed by analyzing
the low-energy conformations in this regime. The higher the $\epsilon_s$ 
value, the larger the average number of monomers being docked 
at the surface, but a clear cut from the compact adsorbed 
conformations does not exist. This difference in the wetting 
transition for $N=13$ and $N=20$ might be due to the fact that 
the most compact conformation for $N=13$ is an almost perfect 
icosahedron~\cite{stefan}. ``Almost'' because the length scales of covalent
bonds and intermonomeric Lennard-Jones interaction differ
slightly.

We also searched for low-energy states with a modified 
LJ energy minimum distance shifted to unity and indeed 
found perfectly icosahedral morphologies. This additionally 
stabilizes the polymer conformation and is already known 
from studies of atomic clusters. The smallest Mackay
icosahedron \cite{mackay} with characteristic 5-fold symmetry is formed
by 13 atoms. Larger perfect icosahedra also require a ``magic''
number (55, 147, 309,\ldots) of atoms. This holds also true for
crystals of elastic polymers\cite{stefan}. Thus, it might be worthwhile 
to also study the wetting transition for other chain lengths 
in order to be able to predict a trend for longer chains, which 
is not possible only knowing the behavior for the two chain 
lengths investigated in our study. The parameters of the low-temperatures 
pseudophase transitions for the 13mer and the 20mer are summarized in Table \ref{tab:1}.
\begin{center}
\begin{table}
\caption{Surface Attraction Strength $\epsilon_s$ for all Low-temperature Transitions for $N=13$ and $N=20$}
\label{tab:1}
\begin{tabular}{lllll}
 \hline
 \hline
$\phantom{x}$ transition & $\phantom{x}$ & $N=13$ & $\phantom{x}$ &  $N=20$ $\phantom{x}$\\
 \hline
\begin{footnotesize}$\phantom{x}$ adsorption transition\end{footnotesize} & $\phantom{x}$ &  $\epsilon_s\approx 0.2$ & $\phantom{x}$ & $\epsilon_s\approx 0.2$\\
\begin{footnotesize}$\phantom{x}$ transition AC $\leftrightarrow$ AC2a\end{footnotesize} & $\phantom{x}$ &  $\epsilon_s\approx 0.5$ & $\phantom{x}$ & ~~~~~--\\
\begin{footnotesize}$\phantom{x}$ transition AC2a $\leftrightarrow$ AC2b\end{footnotesize} & $\phantom{x}$ &  $\epsilon_s\approx 0.9$ & $\phantom{x}$ & $\epsilon_s\approx 1.7$\\
\begin{footnotesize}$\phantom{x}$ layering transition AC2b $\leftrightarrow$ AC1\end{footnotesize} & $\phantom{x}$ &  $\epsilon_s\approx 2.8$ &$\phantom{x}$ &  $\epsilon_s\approx 3.4$\\
 \hline
 \hline
\end{tabular}\end{table}\end{center}

Raising the temperature above the freezing temperature, 
polymers from adsorbed and rather compact conformations
that look like glocular, unstructured drops on the surface. This 
pseudophase is called a \textit{surface-attached globule} (AG) phase.
It has been first conjectured from short exact enumeration studies 
of 2D polymers in poor solvent~\cite{kumar1} but was also found in lattice-polymer
studies \cite{michael,prellberg}. 
\begin{figure}
 		\includegraphics[width=8.7cm]{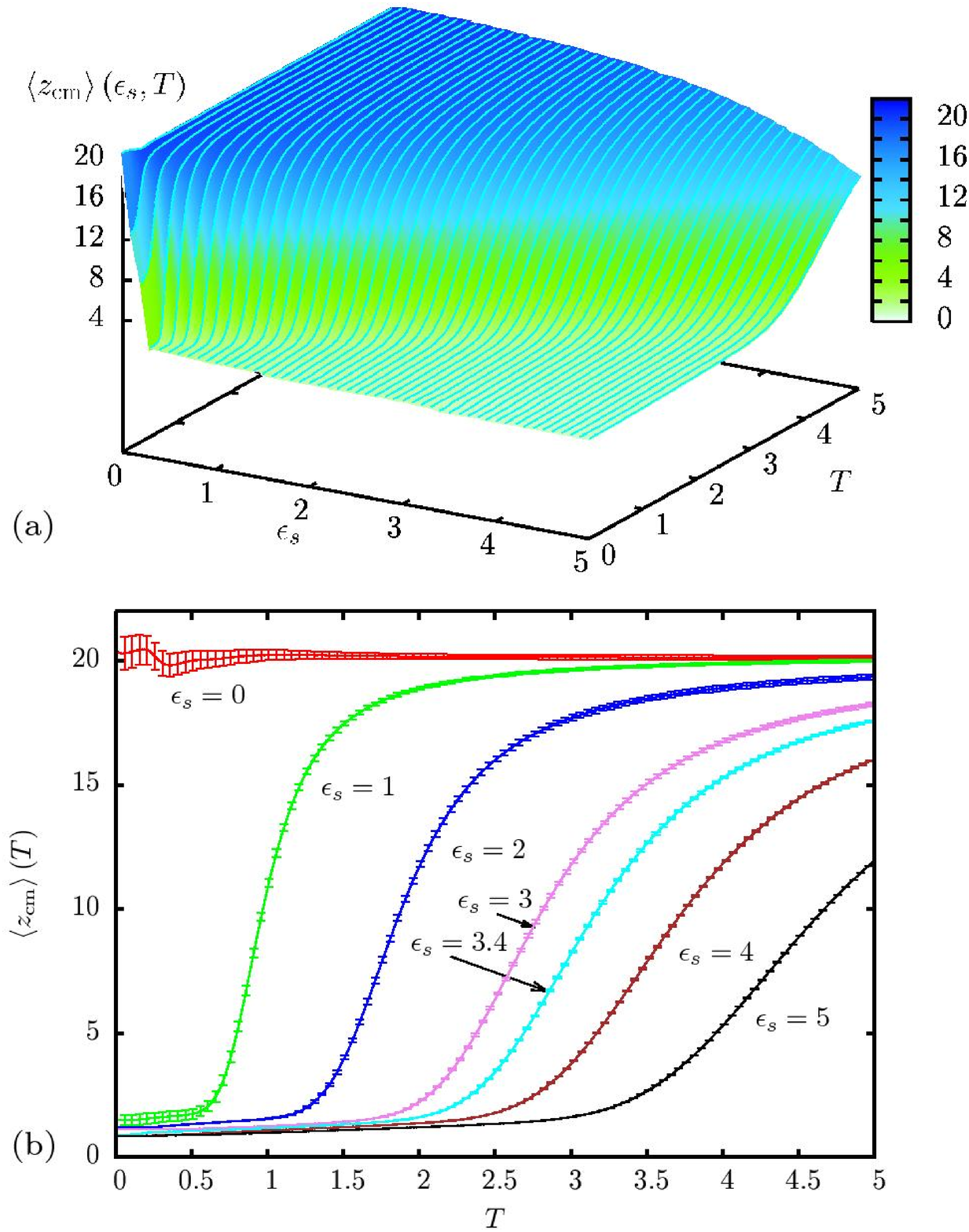}
\caption{\label{fig:zcm3d} (a) $\left\langle z_{\rm cm}\right\rangle $ of the 20mer. (b) $\left\langle z_{\rm cm}\right\rangle $ for various values of 
the parameter $\epsilon_s$. }
\end{figure}
At even higher temperatures, two scenarios
can be distinguished in dependence of the relative strengths of
monomer-monomer and monomer-substrate interactions. In
the first case, the polymer first desorbs from the surface [from 
AG to the desorbed globular (DG) bulk phase] and disentangles
at even higher tempera\-tures [from DG to the desorbed expanded 
bulk phase (DE)]. In the latter case, the polymer expands while 
it is still on the surface (from AG to AE2) and desorbs at higher 
tempera\-tures (from AE2 to DE). Due to the higher relative 
number of monomer-monomer contacts in compact bulk 
conformations of longer chains, the $\theta$-temperature increases with 
$N$. The same holds true for the surface attraction strength, $\epsilon_s$, 
associated with the layering transition. 

It is clear that the structural behavior of the studied small
chains is affected by finite-size effects, in particular in the
compact pseudophases. As long as surface effects are as
influential as volume effects, the shapes of compact adsorbed
(but also of compact desorbed\cite{stefan}) conformations differ noticeably
for polymers with different but small lengths, and a precise
classification is difficult. However, for longer chains, filmlike
(AC1) and semispherical conformations (AC2), as well as
surface-attached droplets (AG), will dominate the respective
phases. Currently, the simulation of longer chains, aiming at
the identification of all conformational phases and the quantitative
analysis in the thermodynamic limit, is too challenging. 
Thus, a more detailed classification within the compact phases is left for future work. 
\subsection{Adsorption Parameters}
\begin{figure}
 		\includegraphics[width=8.8cm]{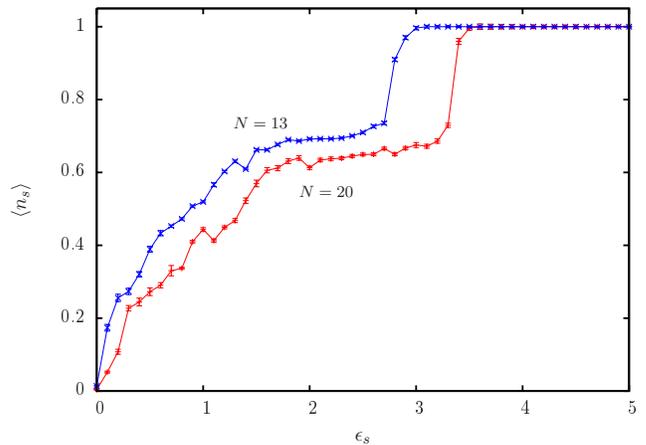}
\caption{\label{fig:2} Mean number of surface contacts $\left\langle n_s\right\rangle$ vs.~surface attraction 
strength $\epsilon_s$ for the 13mer and the 20mer at $T=0.001$.}
\end{figure}

\begin{figure}
 		\includegraphics[width=8.7cm]{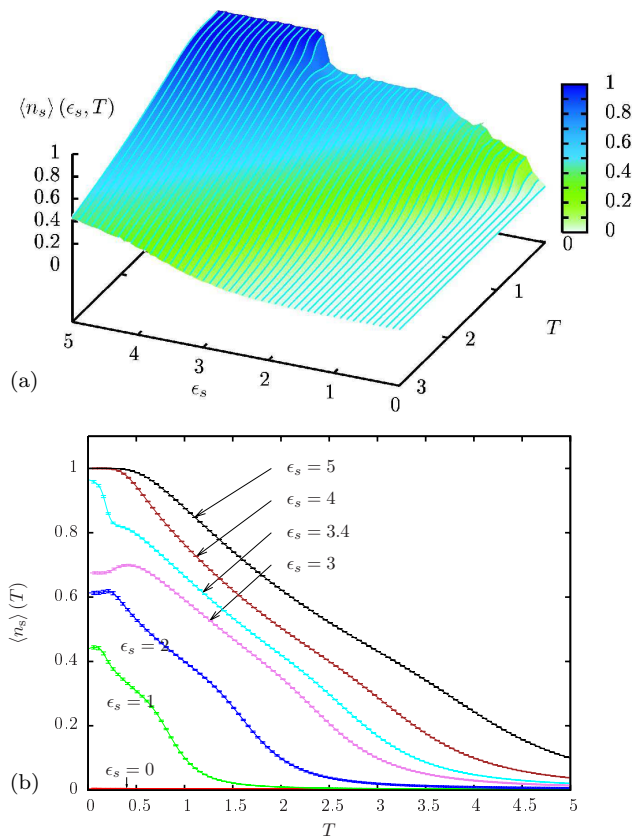}
\caption{\label{fig:docking3d} (a) $\left\langle n_s\right\rangle $ of the 20mer. (b)  $\left\langle n_s\right\rangle $ for different values of $\epsilon_s$. }
\end{figure}

The adsorption transition can 
be discussed best when looking at the distance of the center-of-mass 
of the polymer to the surface (Figure \ref{fig:zcm3d}) and the mean 
number of surface contacts (Figures \ref{fig:2} and \ref{fig:docking3d}). As can be seen 
in Figure \ref{fig:zcm3d}, for large temperatures and small values of $\epsilon_s$, the 
polymer can move freely within the simulation box and the 
influence of the substrate is purely steric. Thus, the average 
center-of-mass distance $\left\langle z_{\rm cm}\right\rangle $ of the polymer above the surface 
is just half the height of the simulation box. On the other hand, 
for high $\epsilon_s$ values and low temperatures, the polymer favors
surface contacts and the average center-of-mass distance
converges to $\left\langle z_{\rm cm}\right\rangle \approx 0.858$, corresponding to the minimum-energy
distance of the surface attraction potential for single-layer 
structures, and slightly larger valuesr for double-layer and globular structures. 

One clearly identifies a quite sharp adsorption transition that 
divides the projection of $\left\langle z_{\rm cm}\right\rangle $ in Figure \ref{fig:zcm3d}a into an adsorbed 
(bright/green) regime and a desorbed (dark/blue) regime. This 
transition appears as a straight line in the phase diagram and is
parametrized by $\epsilon_s\propto T$. Intuitively, this makes sense since at 
higher $T$ the stronger Brownian fluctuations are more likely to 
overcome the surface attraction.

Consistently with our above discussion, $\left\langle z_{\rm cm}\right\rangle $ and ${\rm d}\!\left<z_{\rm cm}\right>/{\rm d}T$ 
also reveal the low-temperature transitions between the adsorbed
phases AC2a, AC2b, and AC1. for the detailed discussion of
these adsorbed phases, we concentrate ourselves on the mean
number of surface contacts $\left\langle n_s\right\rangle$ (Figs.~\ref{fig:2}, \ref{fig:docking3d}). 

Unlike in simple-cubic lattice studies, where one finds $\left\langle n_s\right\rangle \approx1$ 
for an $l$-layer structure~\cite{prellberg}, we find for double layer structures
$\left\langle n_s\right\rangle >1/2$. The reason is that most compact multilayer structures
are cuboids on the lattice, whereas in our off-lattice study, 
``layered'' conformations correspond to semispherical shapes,
where, for optimization of the surface of the compact shape,
the surface layer contains more monomers than the upper layer. 
Since this only regards the outer part of the layers, the difference 
is more pronounced the shorter the chain is. 

In Figure \ref{fig:2}, $\left\langle n_s\right\rangle $ is shown as a function of $\epsilon_s$ for small 
temperatures. $\left\langle n_s\right\rangle $ is a useful quantity to identify layering effects. 
Starting at high $\epsilon_s$, for both chain lengths first $\left\langle n_s\right\rangle \approx1$ until at 
the layering transition, $\left\langle n_s\right\rangle $ jumps to $\left\langle n_s\right\rangle \approx 0.69$ for $N=13$ 
and to $\left\langle n_s\right\rangle \approx 0.65$ for $N=20$.

Further jumps corresponding to further layering transitions 
are not observed. Instead, what follows is a plateau regime where 
the relative amount of monomers that cover the surface is rather 
constant. When the double-layer structure gets unstable at lower 
$\epsilon_s$,  $\left\langle n_s\right\rangle $ starts to decrease again. The conformations in AC and 
AC2a thus do not exhibit a pronounced number of surface
contacts, and $\left\langle n_s\right\rangle $ varies with $\epsilon_s$. Near $\epsilon_s\approx0.2$, where the 
polymer desorbs, $\left\langle n_s\right\rangle $ converges rapidly to $\left\langle n_s\right\rangle =0$ as $\epsilon_s\rightarrow 0$.
To conclude, the single- to double-layer ``layering transition''
is a topological transition from 2D to 3D polymer conformations
adsorbed at the substrate. The solvent-exposed part of the 
adsorbed compact polymer structure, which is not in direct
contact with the substrate, reduces under poor solvent conditions
the contact surface to the solvent. Because of the larger number
of degrees of freedom for the off-lattice polymer, layered
structures are not favored in this case. Thus, higher-order
layering transitions are not identified in our analyses (which in
part is also due to the short lengths of the chains) but are also
not expected in pronounced form.

The observable left to discuss is the mean number of intrinsic 
contacts. It behaves very much like the radius of gyration, such 
that the projection of $\left\langle n_m\right\rangle $ onto the $T$-$\epsilon_s$-plane is divided into 
a compact regime comprising AC, AG, AC2a, AC2b, DC, and DG 
and a regime of less compact conformations. This nicely 
confirms the results already obtained. 
\subsection{The Pseudophase Diagram}
To summarize all the 
informations gained from the different observables, we construct 
the approximate boundaries of different regimes in the $T$-$\epsilon_s$
plane. The pseudophase diagram was already displayed in Figure \ref{fig:4} 
where the different pseudophases are denoted by the abbreviations 
explained in the previous subsection. 

The pseudophases found are (for selected representative 
conformations see Figure \ref{fig:5}):
\begin{itemize}
\vspace{-0.2cm}
\item \textbf{DE} (desorbed expanded): Random-coil phase of the quasifree desorbed polymer.
 \vspace{-0.2cm}
\item \textbf{DG} (desorbed globular): Globular phase of the desorbed chain.
 \vspace{-0.2cm}
\item \textbf{DC} (desorbed compact): Maximally compact, spherically shaped crystalline structures dominate this desorption phase 
below the freezing-transition temperatue.
 \vspace{-0.2cm}
\item \textbf{AE1} (adsorbed expanded single layer): Adsorbed phase of expanded, rather planar but little compact random-coil conformations. 
 \vspace{-0.2cm}
\item \textbf{AE2} (adsorbed expanded 3D conformations): Adsorbed, unstructured random-coil-like expanded conformations with
typically more than half of the monomers in contact with the attractive substrate are favored in this pseudophase. 
 \vspace{-0.2cm}
 \item \textbf{AC1} (adsorbed compact single layer): Phase of adsorbed circularly compact filmlike conformations. 
 \vspace{-0.2cm}
 \item \textbf{AG} (adsorbed globular 3D conformations): Representative conformations are surface-attached globular conformations and
look like drops on the surface.
 \vspace{-0.2cm}
 \item \textbf{AC2a} (adsorbed compact 3D conformations): Compact, semispherically shaped crystalline conformations dominate in this subphase. 
 \vspace{-0.2cm}
 \item \textbf{AC2b} (adsorbed compact double layers): Subphase of adsorbed, compact double-layer conformations. The occupation 
of the surface layer is slightly larger than that of the other layer. 
\end{itemize}
AC2a and AC2b are subphases in the regime of the phase 
diagram, where adsorbed compact and topologically three-dimensional
conformations are dominant. Since pronounced
layering transitions as observed in lattice-polymer studies are
not expected here, the discriminations of AC2a and AC2b is
likely to be irrelevant in the thermodynamic limit. However,
AC2a,b differ qualitatively from the phase AC1 of topologically
2D polymer films and we thus expect that the transition between
filmlike (AC1) and semispherical conformations (AC2) is of
thermodynamic relevance.

For $N=13$, where the maximally compact conformation is 
more stable due to the high symmetry of the icosahedral 
structure, we found an additional subphase:
\begin{itemize}
 \item \textbf{AC} (adsorbed icosahedral compact conformations): Like DC, but polymers typically are in touch with the surface. 
As a clear individual subphase only observed for $N=13$.
\end{itemize}

\begin{figure}
 		\includegraphics[width=8.7cm]{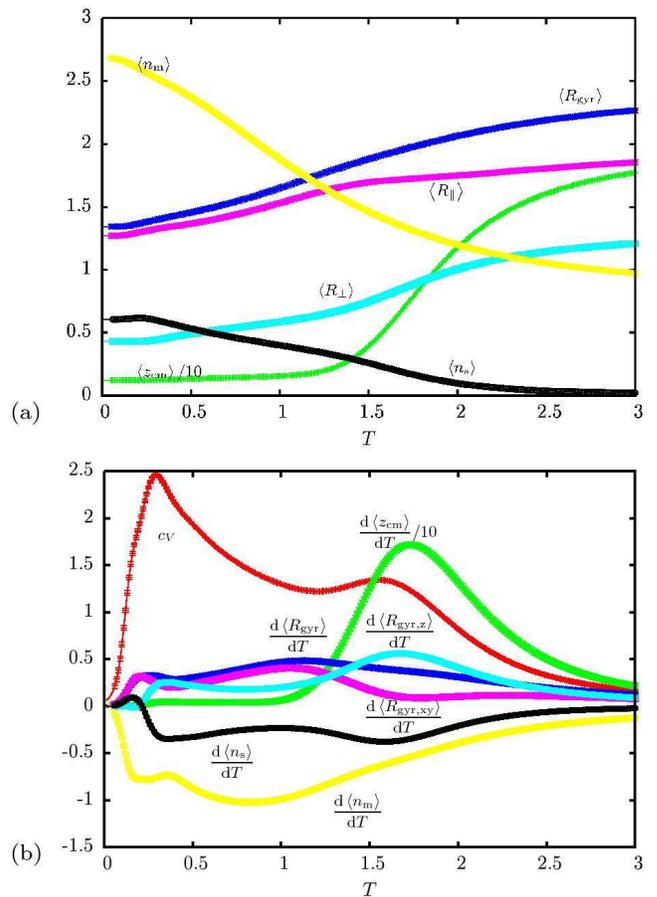}
\caption{\label{fig:6} (a) Temperature dependence of several observables for $\epsilon_s=2$ and $N=20$. 
(b) The derivative with respect to $T$ of the same quantities.}
\end{figure}

The transition lines in the pseudophase diagram (Figure \ref{fig:4}) 
represent the best compromise of all quantities analyzed
separately in our study. Only in the thermodynamic limit of 
infinitely long chains are most of the identified pseudophase
transitions expected to occur at sharp values of the parameters
$\epsilon_s$ and $T$ for all observables. For finitely long chains, the 
transition lines still vary with chain length $N$ and are not well 
defined due to the broad peaks that are slightly different for 
different observables (see Figure \ref{fig:6}). Taking that into account,
the pseudophase diagram gives a good qualitative overview of
the behavior of polymers near attractive substrates in dependence
of environmental parameters such as solvent quality and
temperature. The locations of the phase boundaries should be
considered as rough guidelines.

\section{\label{sec:lattice}Comparison with Lattice Results}
Finally, we would like to compare our results discussed here
with those obtained from a similar model on a simple-cubic (sc) lattice~\citep{michael,mbhetero}.

\begin{figure}
 		\includegraphics[width=8.cm]{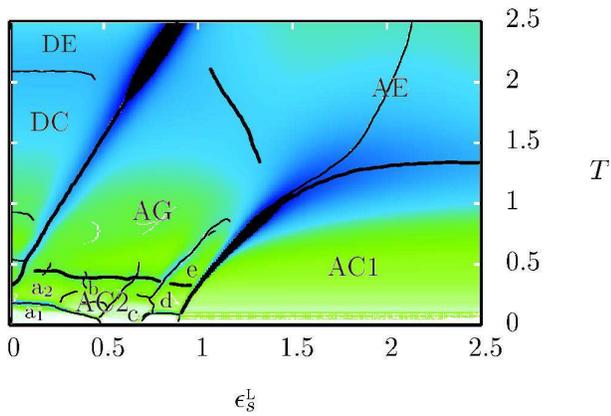}
\caption{\label{fig:pd179} Pseudophase diagram of a lattice polymer with 179 
monomers\cite{michael}, parametrized by the surface attraction strength, $\epsilon_s^{\rm L}$, and
temperature, $T$. The color encodes the specific-heat profile; the 
darker the color, the larger its value.}
\end{figure}

The lattice polymer is modeled as a nongrafted interacting 
self-avoiding walk confined between two infinitely extended 
parallel planar walls. One wall is short-range attractive, and the 
other is purely sterical and prevents the polymer from escaping. 
The energy of the system is given by
\begin{equation}
 E_{\rm L}\left( n_s^{\rm L},n_m^{\rm L}\right) = -\epsilon_s^{\rm L} n_s^{\rm L} - \epsilon_m^{\rm L} n_m^{\rm L},
\end{equation}
where $n_s^{\rm L}$ is the number of nearest-neighbor monomer-substrate 
contacts, $n_m^{\rm L}$ is the number of nearest-neighbor but nonadjacent 
monomer-monomer contacts, and $\epsilon_s^{\rm L}$ and $\epsilon_m^{\rm L}$ are the respective 
contact energy scales. In refs \cite{michael} and \cite{mbhetero} the contact density $g_{n_s,n_m}$
was directly sampled by means of the contact-density chain-growth 
method, which is an extension of the multicanonical 
chain-growth method~\citep{chaingrowth,prellbergflat}. The pseudophase diagram, parameterized 
by temperature and monomer-monomer interaction 
strengh was discussed mainly using the specific-heat profile. 
For a review see ref \citep{WJbook}. The surface-monomer attraction 
strength was fixed. With the contact density, the specific-heat
profile can be calculated for fixed monomer-monomer interaction 
$\epsilon_m^{\rm L}=1$, while varying the surface attraction 
parameter $\epsilon_s^{\rm L}$ as it was done in the present off-lattice study. The 
resulting pseudophase diagram is depicted for a 179mer in 
Figure \ref{fig:pd179}, where the parametrization chosen in ref \citep{michael} was
rescaled in order to allow for a more direct comparison with
the results of our off-lattice study. Denoting the energy and 
temperature from ref \citep{michael} with $E'$ and $T'$, respectively, the rescaling works as follows:

\begin{equation}
\begin{array}{l}
 \dfrac{E'}{T'}=\dfrac{E_{\rm L}}{T} \quad \Leftrightarrow \quad \dfrac{n_s^{\rm L}+s n_m^{\rm L}}{T'}=
\dfrac{\epsilon_s^{\rm L} n_s^{\rm L}+ n_m^{\rm L}}{T}\\
\\
\Leftrightarrow T=\dfrac{T'}{s}\quad \wedge\quad \epsilon_s^{\rm L}=\dfrac{1}{s}.
\end{array}
\end{equation}
Here, $s=\epsilon_m^{\rm L}/\epsilon_s^{\rm L}$ is the ratio of energy scales of intrinsic and
surface contacts as introduced in ref \cite{michael}. It can be interpreted as
a reciprocal solvent-quality parameter.

Certain similarities between the off-lattice (Figure \ref{fig:4}) and the
lattice pseudophase diagram (Figure \ref{fig:pd179}) are obvious. For 
instance, the adsorption transition line is parametrized in both 
models by $\epsilon_s\propto T$. Different, however, is not only the slope,
which depends on the system's geometry and energy scales,
but also for the off-lattice model the extrapolation of the 
transition line seems to go through the origin $\epsilon_s=0$ and $T=0$, 
whereas there is an offset observed in the lattice-system 
analsis such that the extrapolated transition line roughly crosses 
$\epsilon_s^{\rm L}=0.4$ and $T=0$. We speculate that this might be due to the 
intrinsic cuboidal structure of the polymer conformation on
the sc lattice that possess planar surfaces at low temperatures 
even in bulk. Unlike for off-lattice models, where a compact 
polymer attains a spherical shape, such a cuboidal conformation 
is likely to dock at a substrate without substantial conformational
rearrangements. Here lies an important difference between lattice 
and off-lattice models. The off-lattice model provides, for 
sufficiently small surface attraction strengths, a competition 
between most compact spherical conformations that do not 
possess planar regions on the polymer surface and less compact 
conformations with planar regions that allow for more surface 
contacts but reduce the number of intrinsic contacts.

This also explains why a transition like the one observed for 
$N=13$ between AC and AC2a, the wetting transition, is more 
difficult to observe in adsorption studies on regular lattices.
On the other hand, AC2 conformations at low $T$ and for $\epsilon_s$ between 
the adsorption and the single-double layering transitions can 
be observed in both models. Similarly in both models, there
exists the AG pseudophase of surface-attached globules. 

Whereas for the off-lattice model, apart from the wetting 
transition, there is only the transition from AC2a (semispherically 
shaped) to AC2b (double-layer structures), on the lattice 
AC2 comprises of a zoo of subphase transitions. These are higher-order 
layering transitions. Decreasing the surface attraction at
low temperature, layer after layer is added until the number of 
layers is the same as in the most compact conformation. A lattice 
polymer has no other choice than forming layers in this regime. 
The layering transition from AC1 to AC2 is very sharp in both 
models. Also the shape of the transition region from topologically 
2D adsorbed to 3D adsorbed conformations looks very 
similar. Interestingly, the $\epsilon_s^{\rm L}/\epsilon_m^{\rm L}$-ratio predicted for this transition 
in ref \cite{prellberg} agrees quite well with that observed in our off-lattice
study. For low-energy conformations, it is argued that $l^{3/2}=(1-\epsilon_s^{\rm L}/\epsilon_m^{\rm L})N^{1/2}$ 
on the square lattice. With $l=1.5$ and $N=179$ 
this gives $\epsilon_s^{\rm L}/\epsilon_m^{\rm L}=0.914$ for the single- to double-layer transition, 
which is confirmed by Fig.~\ref{fig:pd179}. We re-expressed this argument 
for a triangular lattice, which describes the low-temperature 
conformations of our off-lattice model better, and obtain $l^{3/2}=2(3-\epsilon_s/\epsilon_m)N^{1/2}/3$ 
for low-energy configurations. This yields 
the larger ratio $\epsilon_s/\epsilon_m=2.235\,(2.384)$ for $N=13\,(20)$, which is
due to the higher coordination number of this geometry. It is in
good agreement with our simulation results for this transition.
The higher coordination number also causes the different slopes 
of the respective adsorption transitions.

To summarize, we conclude that, in particular, the high-temperature 
pseudophases, DE, DC/DG, AG, and AE, nicely 
corres\-pond to each other in both models. Noticeable qualitative
deviations occur, as expected, in those regions of the pseudophase
diagram where compact conformations are dominant. 
\section{\label{sec:summary}Summary}
In this paper, we have constructed the pseudophase diagram 
of thermodynamic conformational phases of a single semiflexible 
homopolymer near an attractive substrate in dependence 
of the external parameters surface attraction strength and temperature. 

For two polymer chains with $N=13$ and $N=20$ monomers, 
respectively, the canonical expectation values of several energetical 
and structural quantities and their thermal fluctuations 
were measured in multicanonical computer simulations over a 
broad range of surface attraction strengths and temperatures.
Conformational phases and phase boundaries and their location
in the pseudophase diagram were identified in precise analyses
of structural fluctuations and suitable adsorption parameters.

Although the computational expense to accurately explore 
such a broad parameter range restricted us to investigate rather 
short chains, we identified the following conformational
pseudophases and pseudophase transitions:
\begin{itemize}
 \item Crystalline structures in the regimes of compact phases. We
identified maximally compact desorbed conformations in bulk
(DC) or adsorbed at the substrate (AC), semispherical compact 
conformations (AC2a) that are distorted by the surface but 
not layered, double-layer conformation (AC2b), and single-layer 
conformations (AC1). 
 \vspace{-0.2cm}
 \item Adsorbed conformations in the globular and expanded
(random-coil) phases. Here, three conformational pseudophases 
were distinguished: unstructured 3D surface-attached globular 
conformations (AG), expanded dissolved but planar adsorbed
conformations (AE1), and 3D expanded random-coil-like adsorbed conformations (AE2).
 \vspace{-0.2cm}
 \item Desorbed conformations. Compact conformations (DC) are 
separated by the freezing transition from globular conformations 
(DG). At even higher temperatures above the $\theta$-transition, 
random-coil conformations are found (DE).
\end{itemize}

The sharpest pseudophase transition identified is the layering 
transition between single- and double-layer-structures. 
Higher-layer conformations were not found for these short chains. 
Unlike in recent studies on a simple-cubic lattice, where for 
weak surface attraction and posi\-tive self-attraction, layering 
transitions were observed until a maximally compact cubic 
structure is reached, off-lattice polymers favor maximally 
compact spherical conformation. Thus, we find the expected
differences in the behavior of off-lattice and lattice polymers
in phases, where compact adsorbed conformations dominate.
For the majority of pseudophases, in particular those that are
assumed to be relevant in the thermodynamic limit, we find,
however, a nice qualitative coincidence. This similarity demonstrates 
the ability of such simple coarse-grained models to 
capture the general adsorption behavior of polymers near 
attractive surfaces. 
The increasing experimental and technological capabilities should allow not only for the experimental 
verification of the described thermodynamic phases but also for 
a detection of the pseudophases of finite polymers. Since 
polymers are naturally of finite length, this problem is one of real physical interest. 

\vspace*{0.5cm}
\section{Acknowledgements}

This work is partially supported by the DFG (German Science 
Foundation) under Grant Nos.\ JA \mbox{483/24-1/2} and the Leipzig 
Graduate School of Excellence ``BuildMoNa''.
Support by a NIC supercomputer time grant (No.~hlz11) of the 
Forschungszentrum J{\"u}lich is acknowledged.
\vspace*{0.5cm}
\end{document}